\documentclass[sigconf]{acmart}

\usepackage{booktabs} % For formal tables
\usepackage[inline]{enumitem} % For inline lists

\newcommand{\urlfootnote}[1]{\footnote{\url{#1} (Accessed: \today)}}
\newcommand{\evalZero}{\textbf{\sffamily X}}
\newcommand{\evalOne}{\textbf{!}}
\newcommand{\evalTwo}{\checkmark}

\newcommand{\evalCatOne}{\begin{tabular}[c]{@{}c@{}}\cite{Dinh2017,Sukhwani2018,Androulaki2018,Baliga2018,Thakkar2018,Gupta2018,Sharma2018}\\ \cite{Javaid2019,Kuzlu2019,Nguyen2019,Foschini2020,Nakaike2020,Bergman2020}\end{tabular}}
\newcommand{\evalCatTwo}{\cite{Inagaki2019,Androulaki2019}}
\newcommand{\evalCatThree}{\cite{Pongnumkul2017,Nasir2018,Takeshi2018,Hao2018,Gorenflo2019,Wang2019}}
\newcommand{\evalCatFour}{\cite{Shalaby2020}}
\newcommand{\evalCatFive}{\cite{Nguyen2019b,Wang2020,Yuan2020,Xu2021}}
\newcommand{\evalCatSix}{\cite{Sukhwani2017}}

\setcopyright{acmcopyright}
\acmDOI{xx.xxx/xxx_x}

\acmISBN{978-1-4503-8713-2/22/04}

\acmConference[SAC'22]{ACM SAC Conference}{April 25 –April 29, 2022}{Brno, Czech Republic}
\acmYear{2022}
\copyrightyear{2022}

\acmArticle{4}
\acmPrice{15.00}

\begin{document}
\title{Porting a benchmark with a classic workload to blockchain: TPC-C~on~Hyperledger~Fabric}
% \titlenote{Produces the permission block, and copyright information}
% \subtitle{Some subtitle if needed}
% \subtitlenote{Some subtitle note}
  
\renewcommand{\shorttitle}{TPC-C on Hyperledger Fabric}

\author{Attila Klenik}
\authornote{Corresponding author}
\affiliation{%
  \institution{Department of Measurement and Information Systems\\Budapest University of Technology and Economics}
  \streetaddress{Magyar Tud\'{o}sok K\"{o}r\'{u}tja 2.}
  \city{Budapest} 
  \state{Hungary} 
  \postcode{1117}
}
\email{attila.klenik@vik.bme.hu}

\author{Imre Kocsis}
% \authornote{Second author note}
\affiliation{%
  \institution{Department of Measurement and Information Systems\\Budapest University of Technology and Economics}
  \streetaddress{Magyar Tud\'{o}sok K\"{o}r\'{u}tja 2.}
  \city{Budapest} 
  \state{Hungary} 
  \postcode{1117}
}
\email{kocsis.imre@vik.bme.hu}

% \renewcommand{\shortauthors}{A. Klenik and I. Kocsis}
\begin{abstract}
Many cross-organization cooperation applications of blockchain-based distributed ledger technologies (DLT) do not aim at innovation at the cooperation pattern level: essentially the same ''business'' is conducted by the parties, but this time without a central party to be trusted with bookkeeping. The migration to DLT is expected to have a negative performance impact, but some DLTs, such as Hyperledger Fabric, are accepted to be much better suited performance-wise to such use cases than others. However, with the somewhat surprising, but ongoing absence of application-level performance benchmarks for DLTs, cross-DLT comparison for ''classic'' workloads and the evaluation of the performance impact of ''blockchainification'' is still ill-supported.
We present the design and Hyperledger Caliper-based open implementation of a full port of the classic TPC-C benchmark to Hyperledger Fabric, complete with a structured approach for transforming the original database schema to a smart contract data model. Initial measurements about the workload characteristics that will affect the design of large-scale performance evaluations are also included.
\end{abstract}



\maketitle

\section{Introduction}
\label{sec:introduction}
Distributed ledgers, predominantly implemented today with blockchain technologies, are able to introduce significant business value to a very wide range of established cross-organizational cooperations \cite{wef}. Distributed, fault- and attack-tolerant consensus over the contents of the ledger necessarily introduces performance inefficiencies in contrast to centralized databases and distributed ones operating under more benign fault assumptions (e.g., only non-malicious, independent, fail-silent node failures). 

However, while unpermissioned cryptocurrency networks began to significantly improve only lately on their historically very low throughput and high latency baseline, \textit{consortial} networks -- bespoke, permissioned, closed networks supporting the business cooperation of a relatively small set of organizations -- have always had the selling point of having \textit{tunable} and \textit{designable} performance. Hyperledger Fabric, one of the leading consortial blockchain frameworks, has been reported to be able to achieve thousands of transactions per second \cite{Androulaki2018} for bitcoin-like asset accounting workloads.

While a growing body of literature documents the various micro-level performance mechanisms in Fabric and tooling is available for running quasi-\textit{microbenchmarks}, there's almost a complete lack of full-fledged \textit{macrobenchmarks}; neither ones focusing on ''blockchain-native'' functionality (as UTXO-style cryptoasset handling) nor full ports of classic workloads are available.

This paper improves on this second aspect of the state of the art by introducing an open implementation of the TPC-C OLTP benchmark\urlfootnote{https://drive.google.com/drive/folders/1X-3X59Lzru8do29pqmOUB292ZrgI-9Wt?usp=sharing}\footnote{\textbf{Permission to disclose the link granted by the track co-chairs}}. In addition to still being a widely used database performance benchmark, for Hyperledger Fabric, TPC-C is especially relevant due to its (physical) warehouse data model and order management transactions -- both reflective of such use cases as blockchain-assisted supply chain management and cooperative digital manufacturing.

At the same time, our port showcases a structured approach towards translating classic, SQL-based data models and their transactions to Fabric. We show that, as expected, moving to Hyperledger Fabric transforms database locking induced delays to \textit{retry-induced delays}. More importantly, as the standard workload includes a high ratio of \textit{logically necessary} read-write overlaps between the transactions, it highly depends on the workload magnitude, block size and block time that whether such a structured translation leads to an acceptable performance result. 

But the broader point of our contribution -- beyond creating missing tooling and presenting some methodological advances -- lies on the positive side of this statement: such a conservative transformation \textit{can} be sufficient. 
Performance-enhancing techniques are known for creating Hyperledger Fabric data models broadly based on bitcoin-style UTXOs; but methodologically applying these for existing applications is not mature yet and presents a much greater challenge.
\section{Assessment of Fabric performance}
\label{sec:assessment-of-fabric-performance}

We give a brief introduction to Fabric and briefly survey its performance evaluation literature. This motivates our primary contribution: an open-source, representative, and widely known macro-benchmark implementation for Fabric.

\subsection{Hyperledger Fabric concepts}

Hyperledger Fabric~\cite{Androulaki2018} is a consortial, permissioned DLT. A Fabric network is operated by \textit{organizations} that each contributes computing resources (nodes) to the network. Fabric client applications (\textit{clients} for short) submit transactions to the nodes, but are not strictly part of the network. Nodes can take the roles of \textit{peers} or \textit{orderers}. 

Peers \textit{execute, validate and commit} transactions, i.e., maintain the ledger and its corresponding versioned, \textit{key-value model} world-state. Peers can only execute (\textit{endorse} -- see below) transactions if the corresponding smart contract (called \textit{chaincode} in Fabric) is deployed on the peer. If not, the peer only validates and commits the transaction effects (i.e., all peers validate and commit).

Orderers form the \textit{ordering service} of the network, responsible for determining a unique and global order for transactions. Orderers perform the batching of transactions into blocks; the ordering approach is ''pluggable'' and can range from a single central machine to true, distributed, byzantine fault tolerance implementations.

A given network of nodes does not equal a single ledger. Fabric employs a so-called \textit{channel} mechanism to allow a subset of nodes (and organizations) to form and maintain a ledger isolated from the rest of the network for privacy. Every operation in the Fabric network happens in the scope of a channel. The rest of the paper ignores channels for the sake of readability and assumes a single channel containing every network node.

\begin{figure}
\includegraphics{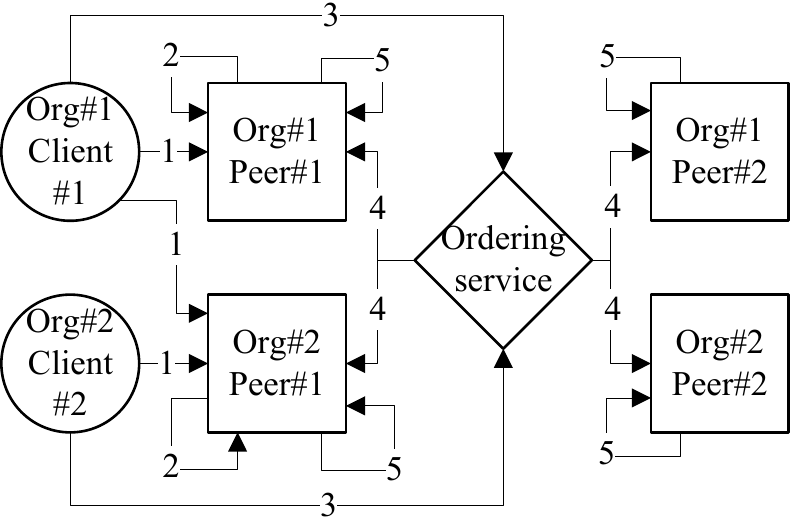}
\caption{Main steps of the Fabric consensus protocol (1-3: transaction-level operations, 4-5: block-level operations)}
\label{fig:fabric-consensus}
\end{figure}

Consensus in Fabric can be outlined as follows (see Fig.~\ref{fig:fabric-consensus}):
\begin{enumerate}
    \item A client sends a \textit{transaction proposal} to one or more peers, operated by one or more organizations (determined by a policy) for so-called \textit{endorsement}.
    
    \item The peers execute the corresponding chaincode based on their \textit{current} blockchain-backed world-state, collecting a versioned set of keys read and to be written as an execution side-effect. This results in a transaction \textit{read-write set}. Writes are \textit{not persisted} at this point yet. Peers sign and send the execution results (''endorsements'') back to the client, including the read-write set.
    
    \item Once the client collected enough endorsements, it packs them into an endorsed \textit{transaction} and sends it to the ordering service for inclusion in a block.
    
    \item The orderer service cuts a new block as soon as one of several pre-configured criteria are met, then \textit{broadcasts} the block to every peer for validation and commit. 
    
    \item If a transaction passes the initial syntax and policy checks, then its read-write set is validated against the current ledger state. As a form of Multiversion Concurrency Control (MVCC), the write values of a transaction are committed if and only if the versions of its read keys are still unchanged in the world-state (including the effects of earlier transactions in the same block). Finally, peers can send notifications about the commit event to subscribed clients.
\end{enumerate}

For a detailed description, please refer to \cite{Androulaki2018}.

\subsection{Fabric performance: state of the art}
Fabric performance is a complex subject, with the existing body of research being dividable into three broad categories.
\begin{itemize}
    \item Evaluating and characterizing the performance of Fabric  \cite{Pongnumkul2017,Baliga2018,Thakkar2018,Nasir2018,Gupta2018,Takeshi2018,Sharma2018,Hao2018,Wang2019,Kuzlu2019,Nguyen2019,Inagaki2019,Nguyen2019b,Androulaki2019,Foschini2020,Wang2020,Shalaby2020,Bergman2020}. Focus falls mainly on the analysis of performance sensitivity to different setups, configurations and workload characteristics. Bottleneck identification and robustness evaluation are also topics of interest.
    \item Performance improvement via various optimization techniques \cite{Thakkar2018,Gorenflo2019,Javaid2019,Nakaike2020}. The proposed improvements target either the steps of the consensus process or the architecture itself.
    \item Formal models of the consensus process \cite{Sukhwani2017,Sukhwani2018,Jiang2020,Yuan2020,Xu2021}. Model parameter identification is usually based on an initial empirical evaluation of the system. Then, subsequent sensitivity analyses with varying configurations are only performed on the model level, reducing the cost of analysis. 
\end{itemize}

Regardless of the aim of the research, designing empirical performance evaluations and reporting their results is a common theme. The Hyperledger Performance and Scalability Working Group (PSWG) released a whitepaper\urlfootnote{https://www.hyperledger.org/learn/publications/blockchain-performance-metrics} about consistently reporting the different aspects of blockchain performance evaluations.

To assess repeatability, we evaluated the level of disclosure of the available results along six dimensions:
\begin{enumerate*}[label=(\roman*)]
    \item hardware environment (HE);
    \item software environment (SE);
    \item network setup and configuration (NS);
    \item test harness setup (TH), i.e., workload generators and data collection;
    \item smart contract specification (SC);
    \item and workload specification (WL).
\end{enumerate*}
The evaluation was permissive; we did not require the disclosure of exact artifacts (e.g., source code of contracts). Even providing a detailed enough description to reproduce the results was deemed sufficient (even if reproduction would require a partial reimplementation of the artifacts).

\begin{table}[]
\begin{tabular}{c||c|c|c|c|c|c}
\textbf{References} & \textbf{HE} & \textbf{SE} & \textbf{NS} & \textbf{TH} & \textbf{SC} & \textbf{WL} \\
\hline \hline

\evalCatOne   & \evalTwo  & \evalTwo  & \evalTwo & \evalTwo  & \evalTwo  & \evalTwo \\
\hline
\evalCatTwo   & \evalTwo  & \evalTwo  & \evalTwo & \evalTwo  & \evalOne  & \evalTwo \\
\hline
\evalCatThree & \evalTwo  & \evalTwo  & \evalTwo & \evalTwo  & \evalTwo  & \evalOne \\
\hline
\evalCatFour  & \evalZero & \evalZero & \evalTwo & \evalZero & \evalTwo  & \evalOne \\
\hline
\evalCatFive  & \evalTwo  & \evalTwo  & \evalTwo & \evalTwo  & \evalZero & \evalOne \\
\hline
\evalCatSix   & \evalTwo  & \evalTwo  & \evalTwo & \evalTwo  & \evalTwo  & \evalZero          
\end{tabular}
\caption{Repeatability evaluation of Fabric performance experiments (\evalZero: missing; \evalOne: insufficient; \evalTwo: sufficient)}
\label{tab:evaluation}
\end{table}

Table \ref{tab:evaluation} summarizes our evaluation. Reporting of aspects (i)--(iii) and (v) were almost consistently well-covered. However, the specifics of workloads (and, to some extent, smart contracts) show a different picture.

When the workload consists of openly available micro- or macro-benchmarks (e.g., provided by BlockBench \cite{Dinh2017}), then all attributes of the workload are certainly unambiguous. Otherwise, a detailed specification of the applied transaction types and the workload(s) composed from them is necessary. The first aspect is largely well-covered; however, at least in the papers using macrobenchmark or macrobenchmark-like workloads, the rate and transaction type composition of workloads did not receive proper attention.

There are two key issues with this state of the art. First, these specifics are vital not only to reproduce results, but also for meaningful comparisons across different studies. Second, the maturation of DLT has reached the point where the comparison of DLT capabilities with such macro-benchmarks as, e.g., the ones specified by the Transaction Processing Council (TPC) is not only meaningful, but even sorely missing for gauging the technological design space for new DLT projects.

%Such details can significantly affects the validity of, for example, behavior models, where model parameters are estimated and used in consecutive analyses. The authors recognize the priority of model descriptions over the details of the empirical evaluation, albeit an openly available standard benchmark implementation would allow for a succinct workload specification.

%The TPC-C benchmark also includes the specification of the workload in details (not necessarily true for every benchmark). Moreover, a single parameter (the number of warehouses) completely describes the characteristics of the workload, making it trivial to report the sometimes neglected workload specification part of performance evaluations.

TPC-C, a widely accepted and used OLTP benchmark, is a good initial target to address this shortcoming, with a clearly defined workload and a single workload scaling parameter. It is not DLT specific -- what can actually be an advantage when such scenarios are considered where existing business cooperations are migrated to a DLT.

To the best of our best knowledge, an open TPC-C implementation that targets Fabric does not exist. Even though the workload generator part of the standard is implemented by multiple frameworks (OLTP-Bench \cite{Difallah2013}, TPCC-UVa \cite{Llanos2006}, HammerDB\urlfootnote{https://www.hammerdb.com/}, BenchmarkSQL\urlfootnote{https://github.com/pgsql-io/benchmarksql}, Emerald Test\urlfootnote{https://gitlab.com/emerald-platform/emerald}), they all target specific databases (or database connector frameworks), relying on SQL support. Accordingly, their transaction profile implementations are not compatible with Fabric; thus, a specific chaincode-based implementation is needed.
\section{TPC-C}
\label{sec:tpc-c}
TPC benchmarks \textit{specify} a detailed set of measurement and experimental requirements and tend to include software artifacts only very conservatively; they follow the principle that vendors should be able to implement the ''system under test'' without major restrictions (which could artificially hamstring their technology). The TPC-C performance benchmark specifies\urlfootnote{http://tpc.org/tpc_documents_current_versions/pdf/tpc-c_v5.11.0.pdf} an online transaction processing (OLTP) workload that simulates the typical activity of a wholesale supplier. The benchmark was approved in 1992 to enable a more complex evaluation than was possible by previous benchmarks, regarding both its database design and transaction specification. The specification describes a mix of read-only and read-write operations (also referred to as transaction profiles) that mimic the complexity and performance characteristics of similar real-life applications. Furthermore, a scalable logical database design details the participating entities, their relationships, and their attributes.

\subsection{Database design and scale}
The core entities are \textit{warehouses}, each having ten sales \textit{districts}. Warehouses maintain \textit{stock} information about the 100k \textit{items} sold by the company. Each district serves 3k \textit{customers} by maintaining \textit{order}, \textit{new order}, \textit{order item}, and \textit{history} information for all customers.

A single warehouse and its corresponding data comprise roughly 500k database entries (in a classic, relational implementation). The specification also states that each warehouse has ten associated remote \textit{terminals} which emit requests on customer activity -- i.e., the workload. Correspondingly, increasing the number of warehouses, the single scaling parameter of TPC-C, scales not only the size of the database, but also the arrival rate of the workload.

\subsection{Transaction profiles}
The standard specifies multiple types of requests (i.e., transaction profiles) to mimic typical wholesale supplier operations. Different profiles exhibit different characteristics regarding their data access pattern, execution frequency, and timing constraints. TPC-C defines the following transaction profiles:
\begin{description}
    \item[New Order] transactions are mid-weight, read-write, and high-frequency requests that enter a complete customer order atomically in a single transaction.
    \item[Payment] transactions are light-weight, read-write, and high-frequency requests that update a customer's balance and corresponding sales statistics.
    \item[Order Status] transactions are mid-weight, read-only, and low-frequency queries that retrieve information about a customer's last order.
    \item[Delivery] transactions are mid-weight, read-write, and low-frequency requests that process/deliver the ten oldest, not yet delivered orders for a warehouse. Terminals submit delivery requests in a deferred/asynchronous mode, i.e., not waiting for their completion before submitting the next request.
    \item[Stock Level] transactions are heavy, read-only, and low-frequency queries that gather the recently sold items  with unsatisfactory stock levels.
\end{description}

Despite the multiple transaction types of the workload, TPC-C focuses only on a ''business metric'', namely the number of orders processed per minute, denoted as \textit{tpmC}.

\subsection{Terminals}

Each warehouse has ten corresponding terminals that generate transactions according to a weighted distribution of the transaction profiles. Each terminal submits requests sequentially, i.e., in a synchronous manner (except for deferred delivery requests), following a simple customer behavior model graph (CBMG). 

Requests are generated by the following simple cycle:
\begin{enumerate*}[label=(\roman*)]
    \item waiting for input screen display (\textit{menu response time});
    \item waiting for user input (\textit{keying time});
    \item waiting for transaction output (\textit{transaction response time});
    \item and selecting the next transaction type (\textit{think time}).
\end{enumerate*}

The cumulative requests of the independent terminals comprise the overall workload of the System Under Test (SUT): the system handling the terminal requests.
\section{A TPC-C port for Fabric}
\label{sec:tpc-c-port-for-fabric}

This section presents the design decisions behind the two main components of the open-sourced artifacts: the SUT-side TPC-C chaincode, and the client-side TPC-C workload generator. 

\subsection{TPC-C chaincode design}

\begin{figure}
\includegraphics{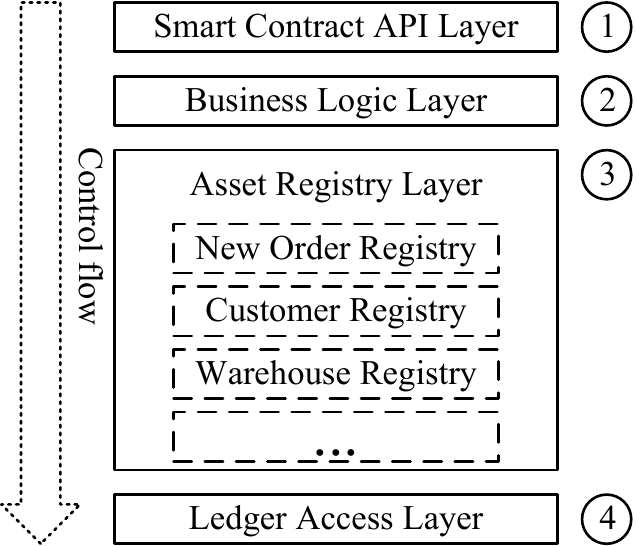}
\caption{TPC-C chaincode layers}
\label{fig:chaincode-design}
\end{figure}

Our TPC-C chaincode follows a layered design (Fig.~\ref{fig:chaincode-design}), separating concerns along different abstraction levels. Correspondingly, each layer is responsible for a specific set of operations, utilizing only the services of the next layer (as detailed in the next sections). 

Even though the chaincode was implemented using Node.JS and the recommended contract design pattern,\urlfootnote{https://hyperledger.github.io/fabric-chaincode-node/release-1.4/api/tutorial-using-contractinterface.html} the loosely coupled, layered design facilitates porting the chaincode to other supported languages, whether the contract pattern is supported (e.g., for Java chaincodes) or not (e.g., for Golang chaincodes). 

\subsubsection{Smart contract API layer}

The first layer is the entry point of the chaincode, receiving control in the appropriate chaincode function targeted by the client. The foremost concern of the layer is to \textit{transform data} between the raw string format (required by the chaincode API) and the ''strongly typed'' representations of the business logic layer. 

Furthermore, the layer also contains domain-independent instrumentation to aid transaction traceability (e.g., logging the beginning and end of transactions), and general error handling mechanisms.

\subsubsection{Business logic layer}

The second layer contains the verbatim implementations of the TPC-C transaction profile specifications. The business logic layer relies only on the CRUD-like (Create, Read, Update, Delete) operations of the registry layer. This design pattern lends an almost pseudo-code-like form to the implementation, substantially simplifying specification conformance checks and requirement traceability. The layer also serves as an instrumentation point for gathering domain-specific metadata, facilitating the identification of workload- and state-dependent performance characteristics.

\subsubsection{Asset registry layer}

The third layer plays the role of a simple object-relational mapping (ORM) framework, hiding the peculiarities of the (ledger-specific) storage implementation. The layer provides CRUD-like operations for each asset type in the form of separate registries (similarly to the now deprecated Hyperledger Composer tool\urlfootnote{https://www.hyperledger.org/use/composer}), while using the domain-agnostic CRUD operations of the ledger access layer.

Fabric employs a key-value store abstraction for persisting the world-state, where keys are arbitrary strings, and values are arbitrary byte blobs. The registries transparently manage the structure of the key space, without affecting the corresponding states (i.e., values) of domain objects. Moreover, the layer also encapsulates complex queries for the business logic layer, e.g., retrieving a customer either by their unique ID or by their last name (depending on the available inputs).

Correspondingly, the registry layer encapsulates the storage access-related design decisions of the implementation, constraining the potential modifications to a single layer when experimenting with other designs and configurations (as expected from a layered service).

\subsubsection{Ledger access layer}

The fourth and last layer encapsulates the API of the chaincode SDK by providing simple CRUD operations on a higher abstraction level than byte blobs. 

More importantly, this layer gives place to the low-level, domain-agnostic instrumentation that tracks the data access statistics of transactions, e.g., the number of read/write operations and the size of values read/written from/to the ledger. Such statistics play a crucial part in understanding the performance characteristics of the Fabric consensus phases.

\subsection{TPC-C chaincode data model}

The standard specifies a relational data model for the entities that needs to be mapped to Fabric's key-value store abstraction. While the mapping is mostly straightforward, special care must be taken in some cases due to the key-value store characteristics and API.

In general, the key space is sharded along the ''database tables'', distinguishing entity types using the database table names as prefixes. Furthermore, the key of each entity is constructed from the concatenation of their primary keys. For example, a warehouse key conforms to the following format: $concat(''WAREHOUSE'', w\_id)$, where $w\_id$ is the sole primary key for warehouse entities, and $concat$ denotes a function of the chaincode API that can construct composite keys amenable to partial lookups.

However, the presented mapping had to be enhanced because:
\begin{enumerate*}[label=(\roman*)]
    \item the chaincode API provides key iterations only in lexicographic order;
    \item and insertion order-based (e.g., oldest, newest entry) key access is necessary for some profiles.
\end{enumerate*}

The key order constraint prompted the following modifications of the key space (transparently handled by the asset registry layer):
\begin{enumerate}
    \item Numeric primary keys are left-padded with zero characters to a fixed length, so that the resulting key $ENTITY\_02$ precedes the key $ENTITY\_11$ in iteration order (making simpler lookups more efficient).
    \item Customer entries are also stored by a secondary, utility key that contains their last names. This enables efficient customer lookups by their last name, as required by payment transactions. Note, that payment transactions are read-write requests, thus CouchDB-based rich queries are not applicable (CouchDB is a backend option for Fabric). Moreover, greedily enumerating every customer to find matches would quickly explode the read-set of the transaction, increasing the number of invalid transactions due to concurrent data access conflicts.
    \item The customer-scoped history entries do not have a primary key specified in the standard, so a client-side timestamp is utilized to create a unique key to avoid overwriting previous history entries. Utilizing client-side timestamp, in general, can open up the chaincode to replay attacks. However, the used timestamp (combined with the client identity) also contributes to the transaction ID, delegating replay attack detection to higher-level Fabric layers. 
    \item The key (and only the key) of order entries contains a flipped order ID, meaning that a monotonically decreasing counter is used instead of an increasing one. This ensures that the last inserted order is iterated first, enhancing the efficiency of order lookup in the order status queries.
\end{enumerate}

Currently, the implementation is tailored to the common key-value store API of Fabric. Creating a variant that utilizes CouchDB-specific rich queries is left for future work.

\subsection{TPC-C workload generator design}

This section presents the design of the TPC-C workload generator and its integration with Hyperledger Caliper.\urlfootnote{https://www.hyperledger.org/use/caliper} 

\subsubsection{Hyperledger Caliper}

Caliper is the official benchmarking tool of the Hyperledger project umbrella. Its modular and scalable design\urlfootnote{https://hyperledger.github.io/caliper/vNext/architecture/} enables generating custom workloads towards large-scale systems. 

Caliper utilizes two types of (micro-)services during benchmarking. An arbitrary number of \textit{worker} services independently generate the workload, while a single \textit{manager} service orchestrates the workers throughout the different rounds of a benchmark.

Worker services are configured with a custom workload module and rate controller for each round. The rate controller determines the \textit{sending rate} of transactions, while the user-implemented workload module dictates the \textit{content} of each transaction. The rate and content aspects of a workload are usually independent for micro-benchmarks (and for many macro-benchmarks), facilitating sensitivity analyses for variances in the workload.

However, TPC-C specifies the scheduling of the workload in detail, not just its content. Accordingly, the presented implementation acts \textit{both as a rate controller and workload module} (Sec.~\ref{subsubsec:workload-module}), resulting in a closed-loop workload generator. 

\subsubsection{Extending Caliper}

TPC-C is a complex performance benchmark with a detailed specification and strict constraints. While Caliper provides many customization opportunities, the authors had to extend its feature set to fulfill the requirements of a full-fledged macro-bechmark, like TPC-C:
\begin{enumerate}
    % \item The benchmark consists of two rounds: first a worker service \textit{loads} the required entries into the SUT; then multiple worker services perform the \textit{execution} of the benchmark according to the specification. Generally, the execution round requires more workers than the load round, which can be performed using a single worker.
    
    % However, Caliper can only utilize the same number of workers for every round. The authors extended the benchmark configuration schema to indicate the \textit{required number of workers} for each round. Combined with the next extension, it became possible to execute both phases with a single Caliper run, easing the measurement design and data harness.
    \item The benchmark configuration schema was extended to indicate the \textit{required number of workers} for each round, allowing using a single worker to load the database, then performing the workload generation with multiple workers, both in the same Caliper run.
    
    % \item The TPC-C workload relies heavily on random number generators for determining entity attributes and transaction input values. Moreover, the standard specifies certain constraints between the random seeds of the execution and the load phase. Furthermore, the chosen seeds must be consistent during the entire benchmark run and across the distributed worker services. Although pre-computing the seeds offline is a viable workaround, it is recommended to use different seeds for different runs to ensure workload diversity.
    
    % However, Caliper does not provide mechanisms for such a direct and distributed state synchronization between worker services. While the task could be accomplished with a shared storage or messaging service between the workload modules of the workers, a much more light-weight, integrated solution is better suited for the goal. 
    
    % The authors extended the workload module interface with an additional \textit{prepare step}, orchestrated by the manager service. The new step allows the workload modules of different rounds to construct arbitrary states, which later will be shared with every worker utilizing the messaging service of Caliper.
    \item The workload module interface was extended with an additional \textit{prepare step}, orchestrated by the manager service. The new step allows the workload modules of different rounds to construct arbitrary states, which later will be shared with every worker utilizing the messaging service of Caliper. The extension allows the dissemination of shared random seeds, as required by the benchmark.
    
    % \item TPC-C also relies on random numbers while populating the database with entries. For example, an order contains ten order lines \textit{in average}, meaning a uniform random number between five and fifteen is used for each order. Thus the total number of order lines can fluctuate significantly between runs, since even a single warehouse setup contains tens of thousands of orders. Correspondingly, the total number of entries to insert cannot be known a priori.
    
    % Caliper, however, only support two driving mode for rounds: run for a specified amount of time, or execute a given number of transactions. It is clear that neither of the modes can support the randomness of TPC-C when inserting entities. The authors implemented a third driving mode where the workload module can \textit{explicitly signal} the worker when it deems the round finished, i.e., when it inserted every required entity.
    \item A transaction number- and time-independent, third driving mode was implemented where the workload module can \textit{explicitly signal} the worker when it deems the round finished, i.e., when it inserted every required entity. Such a driving mode can account for the random number of entries, not conforming to fix transaction numbers or execution time.
\end{enumerate}

\subsubsection{The workload module}
\label{subsubsec:workload-module}

\begin{figure}
\includegraphics{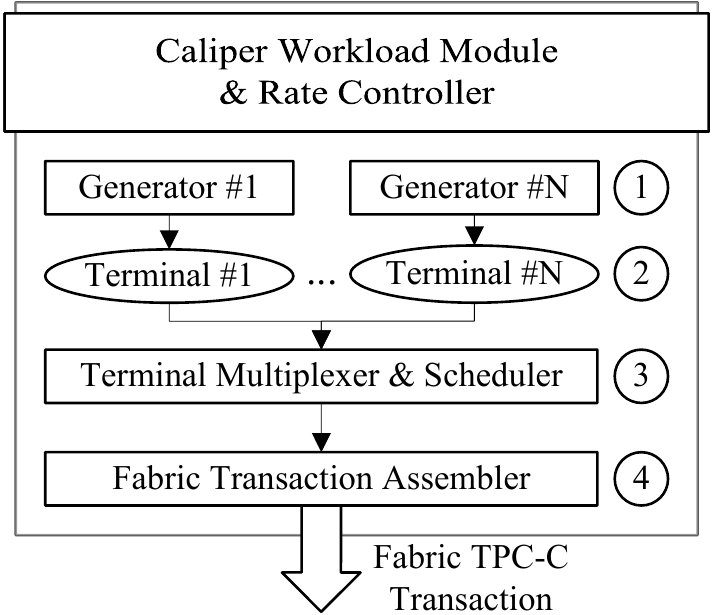}
\caption{Caliper TPC-C workload module components}
\label{fig:workload-generator}
\end{figure}

The workload module for the execution phase (Fig.~\ref{fig:workload-generator}) follows a layered design (the module for the initial database load phase follows a simpler, one-terminal design, thus its description is omitted). Each layer augments the raw TPC-C input data until it becomes a fully configured Fabric transaction. 

The first layer is responsible for generating the raw arguments and the timing constraints for the transaction profiles. The generators are stateless and operate independently upon requests from the terminal layer. The entry and argument generators are based on the implementation of the py-tpcc\urlfootnote{https://github.com/apavlo/py-tpcc} tool (which is ignoring the timing constraints).

The second layer impersonates multiple terminals, each with its own dedicated generators. The terminals implement the sequential behavior, i.e., that new requests are only generated when the previous one is finished (except for deferred delivery requests). Terminals also calculate the scheduling of the next request based on the timing constraints (think time, menu selection time, and keying time) of the previous and next requests. Finally, the terminals push the next request to the next layer.

The third layer ensures that the requests of multiple terminals are scheduled and submitted in order (as calculated by the terminals). Moreover, this multiplexer layer plays the part of a Caliper rate controller. The multiplexer maintains a sorted list of requests, and when Caliper instructs it to ''halt'' until the next transaction submit time, it pops the first/next request and waits until it is time for sending it.

As a naive approach, a Caliper worker service could emulate only a single terminal, greatly simplifying the implementation. However, the sending rate of a single terminal is so low (less than a transaction per second) that the worker service would be underutilized. Moreover, large-scale measurements might require thousands or millions of terminals, which would mean a significant management overhead if each terminal were emulated by a different worker service. 

Correspondingly, the presented workload module implementation (which is instantiated on the Caliper worker level) is capable of emulating multiple terminals. Furthermore, the multiplexer layer is instrumented to gather data about the ''scheduling precision reserve,'' i.e., how well can the rate controller keep up with the calculated scheduling times. The data can help to properly design experiments that result in economical resource utilization while ensuring compliance with the specified scheduling constraints.  

TPC-C request scheduling was SUT-agnostic up until the fourth and last layer, facilitating ports to other SUT types. It is the job of the last layer to encapsulate the arguments into a Fabric-specific transaction, i.e., specify the target nodes, channel, chaincode, and user identity to use. Finally, the assembled transaction is sent using Caliper's Fabric connector.
\section{Experimental evaluation}
\label{sec:experimental-evaluation}

This section presents a preliminary evaluation of the workload generator and smart contract implementation. The evaluation aims at facilitating the design of the scaling and sensitivity analysis aspects of measurement campaigns performed with our TPC-C implementation.

\subsection{Measurement setup}

\begin{figure}
\includegraphics{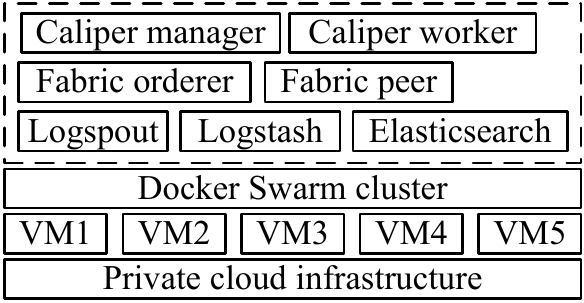}
\caption{Measurement environment setup}
\label{fig:setup}
\end{figure}

Each measurement was performed in a private cloud infrastructure (Fig.~\ref{fig:setup}) on five virtual machines (VM). Each VM was equipped with 4 vCPUs, 8GB memory, and a 40GB HDD, running the Ubuntu 18.04 LTS operating system. The measurement utilized containerized services deployed on a Docker Swarm cluster across the VMs.  

The Fabric network consisted of a single peer and orderer node, both of version 1.4.11. The peer node used GoLevelDB as world-state database, while the orderer cuts new blocks every 100ms (the latest). Other Fabric node configurations retained their default values. 

A Caliper manager service orchestrated a single Caliper worker service, performing both the load and execution round of TPC-C. Each measurement used a scale of one warehouse, but the number of terminals was increased between measurements (from 10 to 100, in steps of ten; and from 100 to 400, in steps of 50). Note, how the \textit{workload is completely specified by only two parameters}. Going by the standard, the number of warehouses alone would suffice; our implementation diverges from that in that we made the warehouse-terminal ratio parameterizable, too.

The data collection part of the test harness utilized the open-source Elastic stack,\urlfootnote{https://www.elastic.co/elastic-stack/} processing logs from the Fabric nodes and the Caliper worker.

\subsection{Workload generation rate and precision}

\begin{figure}
\includegraphics[width=0.8\columnwidth]{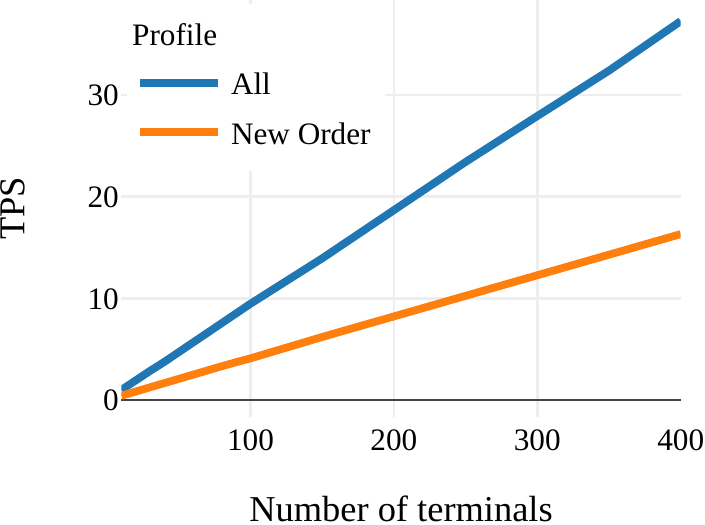}
\caption{Aggregate workload rates of terminals}
\label{fig:terminal-workload-rates}
\end{figure}

\begin{figure}
\includegraphics[width=0.8\columnwidth]{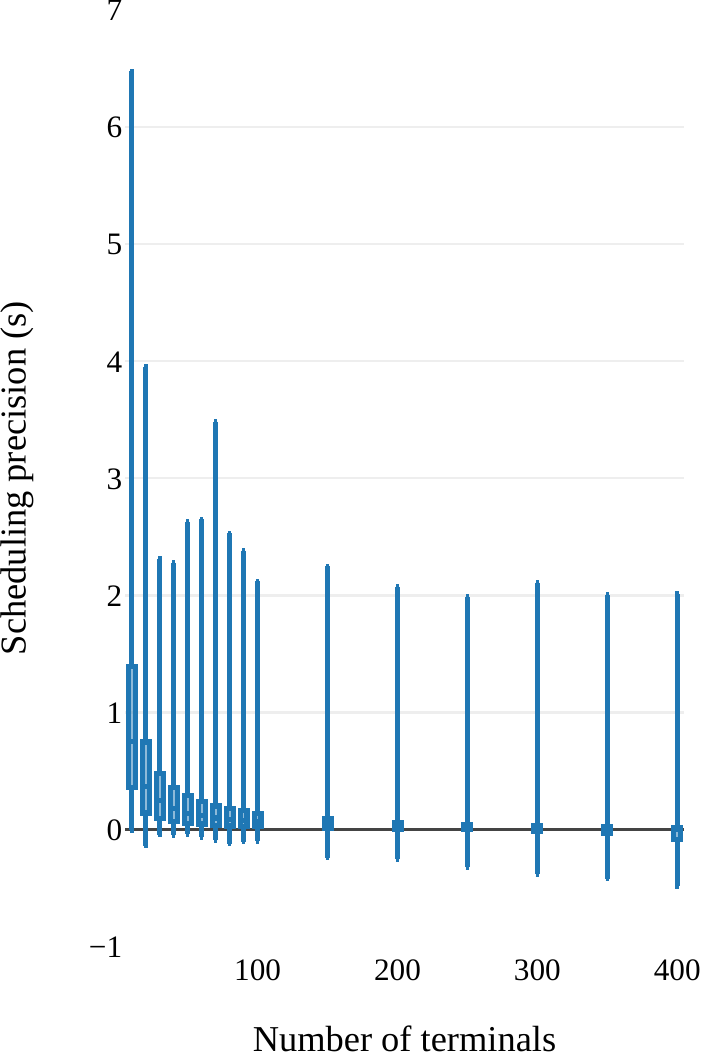}
\caption{Scheduling precision boxplots for one worker}
\label{fig:scheduling-reserves-boxplot}
\end{figure}

Determining the number of required services for workload generation is a crucial step for benchmarking at scale. The timing constraints specified by the standard amount to a rate of approximately one transaction per second (TPS) for ten terminals, and this scales linearly with the number of terminals (Fig.~\ref{fig:terminal-workload-rates}). If a TPC-C measurement would like to hit a certain tpmC throughput, then the required scale (number of warehouses) of the benchmark must be set according to the workload generation capabilities of the terminals. Note, that new order transactions are only approximately 45\% of the total workload.

Once the number of required terminals is determined, the next step is to split the terminals among different services (Caliper workers, in this case). The implementation for Caliper workers utilizes a single-threaded multiplexer for emulating multiple terminals. However, care must be taken not to allocate too many terminals to a single worker to uphold the precision of scheduling. 

The \textit{scheduling precision} is defined as the difference between the time when a transaction is picked as next, and the time when it needs to be actually submitted. For example, the multiplexer pops the next transactions from its queue at time $t_1$, inspects its scheduling data to determine the submission time $t_2$. If the difference $d=t_2-t_1$ is negative, then the timing constraint of the corresponding terminal is violated.

Increasing the number of terminals emulated by a worker has a negative impact on scheduling precision (Fig.~\ref{fig:scheduling-reserves-boxplot}). The scheduling precision can drop below zero in the case of hundreds of terminals. However, even for 400 terminals, the delay/violation of transaction submission never exceeds half a second. Such delay might be acceptable, considering that the timing constraints of TPC-C transactions are in the order of magnitude of seconds.  

\subsection{Error profile of workload scaling}

\begin{figure}
\includegraphics[width=0.8\columnwidth]{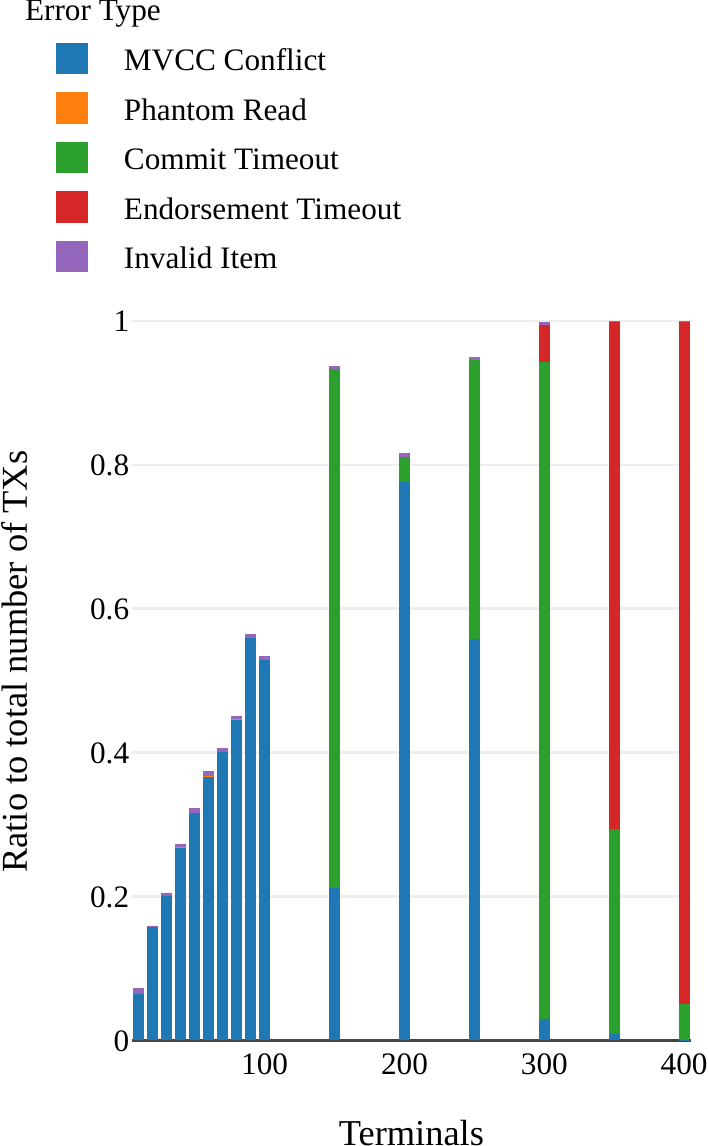}
\caption{Changes in the error profile}
\label{fig:error-ratios}
\end{figure}

In TPC-C, the fixed ratio of terminals to warehouses keeps the concurrent data access situations at a desired level. Traditional databases can resolve such race conditions with locking. For Fabric, concurrent update attempts manifest as MVCC conflicts between two transactions, and the latter one is invalidated, i.e. the client has to submit the request again.

As the number of terminals increase (keeping the number of warehouses fixed), the number of MVCC conflicts increase correspondingly (Fig.~\ref{fig:error-ratios}). Around half of the transactions are invalidated due to MVCC conflicts when the number of terminals is increased to 100. %The measurements beyond 100 terminals show increasing amount of timeout errors from the client's perspective, hiding further MVCC conflicts (in the cases of commit timeouts), and indicating overload issues. 
At even higher terminal numbers, commit timeouts and endorsement timeouts reduce the goodput to effectively zero. The ramifications of these phenomena are twofold.

First, scaling experiments by the benchmark definition (10 terminals per warehouse, no data dependencies across warehouses) will keep the ratio of MVCC conflicts below 10\%, what still can be deemed an acceptable negative impact. In this sense, our approach to port the TPC-C workload proved to be a practically feasible one. However, we also showed that increasing concurrency not only slows down processing -- as with traditional databases -- but \textit{actively wastes} resources; there is a point where other, less straightforward approaches become necessary (as making every single purchasable asset an individual, ownable item and translating ''give me three items'' requests to ''give me those three items'' ones).

Second, commit timeouts can, and endorsement timeouts do represent resource overload situations (predominantly CPU) -- and these happen at relatively low overall load, even for the modest resources we used for error profiling. This does not limit using our smart contracts for either larger-scale benchmarking or configuration sensitivity analysis (as, e.g., adjusting Fabric block time and size), but warrants a further, nuanced analysis of smart contract endorsement resource usage versus Fabric platform resource usage.

%This indicates that our current approach 

%Nevertheless, such variants of the official benchmark are still useful for evaluating the robustness of the chaincode data model design and Fabric performance characteristics. Tuning the ratios of MVCC conflicts using a custom micro-benchmark would require considerable care for the design and implementation, while relying on the existing TPC-C benchmark requires only a single parameter (terminal number) change.

%In general, choosing ''account-like'' entity models to represent, for example, the available stock for a certain item type, will result in concurrent data access upon updates. An UTXO-like model might alleviate such problems, for example, by bookkeeping every individual item separately. However, such choices might require the substantial redesign of the data model. Moreover, it does not eliminate potential conflicts on the individual item level, an additional oracle service might be needed for that.
\section{Conclusion and future work}
\label{sec:conclusion-and-future-work}
Consortial blockchain solutions promise increased performance compared to the available public networks. The permissioned, closed-network nature of such consortial platforms -- Hyperledger Fabric being one of the most mature ones -- enables rigorous capacity planning and performance tuning. However, the performance characterization of the complex Fabric consensus process is still a challenging task. The research related to the performance of Fabric is diverse in its results, and comparisons are difficult to make -- mainly due to the diversity and limited disclosure of the workloads applied. On the industrial side, there are still no practical solutions for \textit{benchmarking} either Hyperledger Fabric, or its contender technologies.

We presented an open implementation of the TPC-C performance benchmark for Fabric. Our artifacts include both a chaincode implementation and the corresponding workload generator. The implementation was evaluated from a perspective that facilitates the design of performance evaluations at scale. As part of our ongoing research, we plan to perform and evaluate such larger-scale experiments and to propose modifications to TPC-C, which incorporate DLT aspects to the benchmark. One straightforward possibility here is transforming the single warehousing organization TPC-C model to a market of suppliers. We also plan to apply the same methodology for porting the IoT-focused benchmark TPCx-IoT to Fabric.

\bibliographystyle{ACM-Reference-Format}
\bibliography{main} 


\begin{thebibliography}{31}

%%% the Web address.

\ifx \showCODEN    \undefined \def \showCODEN     #1{\unskip}     \fi
\ifx \showDOI      \undefined \def \showDOI       #1{#1}\fi
\ifx \showISBNx    \undefined \def \showISBNx     #1{\unskip}     \fi
\ifx \showISBNxiii \undefined \def \showISBNxiii  #1{\unskip}     \fi
\ifx \showISSN     \undefined \def \showISSN      #1{\unskip}     \fi
\ifx \showLCCN     \undefined \def \showLCCN      #1{\unskip}     \fi
\ifx \shownote     \undefined \def \shownote      #1{#1}          \fi
\ifx \showarticletitle \undefined \def \showarticletitle #1{#1}   \fi
\ifx \showURL      \undefined \def \showURL       {\relax}        \fi
% invisible to TeX
\providecommand\bibfield[2]{#2}
\providecommand\bibinfo[2]{#2}
\providecommand\natexlab[1]{#1}
\providecommand\showeprint[2][]{arXiv:#2}

\bibitem[\protect\citeauthoryear{Androulaki et~al\mbox{.}}{Androulaki
  et~al\mbox{.}}{2018}]%
        {Androulaki2018}
\bibfield{author}{\bibinfo{person}{Elli Androulaki} {et~al\mbox{.}}}
  \bibinfo{year}{2018}\natexlab{}.
\newblock \showarticletitle{{Hyperledger Fabric: A Distributed Operating System
  for Permissioned Blockchains}}. In \bibinfo{booktitle}{\emph{Proc. of the
  Thirteenth EuroSys Conf.}} \emph{(\bibinfo{series}{EuroSys '18})}.
  \bibinfo{publisher}{{ACM}}, \bibinfo{address}{{New York, NY, USA}},
  \bibinfo{pages}{30:1--30:15}.
\newblock


\bibitem[\protect\citeauthoryear{Androulaki, {De Caro}, Neugschwandtner, and
  Sorniotti}{Androulaki et~al\mbox{.}}{2019}]%
        {Androulaki2019}
\bibfield{author}{\bibinfo{person}{Elli Androulaki}, \bibinfo{person}{Angelo
  {De Caro}}, \bibinfo{person}{Matthias Neugschwandtner}, {and}
  \bibinfo{person}{Alessandro Sorniotti}.} \bibinfo{year}{2019}\natexlab{}.
\newblock \showarticletitle{{Endorsement in Hyperledger Fabric}}. In
  \bibinfo{booktitle}{\emph{Proceedings - 2nd IEEE Int. Conf. on Blockchain}}.
  \bibinfo{publisher}{{IEEE}}, \bibinfo{address}{{Piscataway, NJ, USA}},
  \bibinfo{pages}{510--519}.
\newblock


\bibitem[\protect\citeauthoryear{Baliga et~al\mbox{.}}{Baliga
  et~al\mbox{.}}{2018}]%
        {Baliga2018}
\bibfield{author}{\bibinfo{person}{Arati Baliga} {et~al\mbox{.}}}
  \bibinfo{year}{2018}\natexlab{}.
\newblock \showarticletitle{{Performance characterization of Hyperledger
  Fabric}}. In \bibinfo{booktitle}{\emph{Crypto Valley Conf. on Blockchain
  Technology}}. \bibinfo{publisher}{{IEEE}}, \bibinfo{address}{{Piscataway, NJ,
  USA}}, \bibinfo{pages}{65--74}.
\newblock


\bibitem[\protect\citeauthoryear{Bergman, Asplund, and Nadjm-Tehrani}{Bergman
  et~al\mbox{.}}{2020}]%
        {Bergman2020}
\bibfield{author}{\bibinfo{person}{Sara Bergman}, \bibinfo{person}{Mikael
  Asplund}, {and} \bibinfo{person}{Simin Nadjm-Tehrani}.}
  \bibinfo{year}{2020}\natexlab{}.
\newblock \showarticletitle{{Permissioned blockchains and distributed
  databases: A performance study}}.
\newblock \bibinfo{journal}{\emph{Concurrency and Computation: Practice and
  Experience}} \bibinfo{volume}{32}, \bibinfo{number}{12}
  (\bibinfo{year}{2020}), \bibinfo{pages}{e5227}.
\newblock


\bibitem[\protect\citeauthoryear{Difallah, Pavlo, Curino, and
  CudreMauroux}{Difallah et~al\mbox{.}}{2013}]%
        {Difallah2013}
\bibfield{author}{\bibinfo{person}{Djellel~Eddine Difallah},
  \bibinfo{person}{Andrew Pavlo}, \bibinfo{person}{Carlo Curino}, {and}
  \bibinfo{person}{Philippe CudreMauroux}.} \bibinfo{year}{2013}\natexlab{}.
\newblock \showarticletitle{{OLTP-bench: An extensible testbed for benchmarking
  relational databases}}.
\newblock \bibinfo{journal}{\emph{Proceedings of the VLDB Endowment}}
  \bibinfo{volume}{7}, \bibinfo{number}{4} (\bibinfo{year}{2013}),
  \bibinfo{pages}{277--288}.
\newblock


\bibitem[\protect\citeauthoryear{Dinh et~al\mbox{.}}{Dinh
  et~al\mbox{.}}{2017}]%
        {Dinh2017}
\bibfield{author}{\bibinfo{person}{Tien Tuan~Anh Dinh} {et~al\mbox{.}}}
  \bibinfo{year}{2017}\natexlab{}.
\newblock \showarticletitle{{BLOCKBENCH: A framework for analyzing private
  blockchains}}. In \bibinfo{booktitle}{\emph{Proc. of the ACM SIGMOD Int.
  Conf. on Management of Data}}, Vol.~\bibinfo{volume}{Part F1277}.
  \bibinfo{publisher}{ACM}, \bibinfo{address}{New York, NY, USA},
  \bibinfo{pages}{1085--1100}.
\newblock
\urldef\tempurl%
\url{https://doi.org/10.1145/3035918.3064033}
\showDOI{\tempurl}


\bibitem[\protect\citeauthoryear{Foschini et~al\mbox{.}}{Foschini
  et~al\mbox{.}}{2020}]%
        {Foschini2020}
\bibfield{author}{\bibinfo{person}{Luca Foschini} {et~al\mbox{.}}}
  \bibinfo{year}{2020}\natexlab{}.
\newblock \showarticletitle{{Hyperledger Fabric Blockchain: Chaincode
  Performance Analysis}}. In \bibinfo{booktitle}{\emph{IEEE Int. Conf. on
  Communications}}, Vol.~\bibinfo{volume}{2020-June}.
  \bibinfo{publisher}{{IEEE}}, \bibinfo{address}{{Piscataway, NJ, USA}},
  \bibinfo{pages}{1--6}.
\newblock


\bibitem[\protect\citeauthoryear{Gorenflo et~al\mbox{.}}{Gorenflo
  et~al\mbox{.}}{2019}]%
        {Gorenflo2019}
\bibfield{author}{\bibinfo{person}{Christian Gorenflo} {et~al\mbox{.}}}
  \bibinfo{year}{2019}\natexlab{}.
\newblock \showarticletitle{{FastFabric: Scaling Hyperledger Fabric to 20,000
  Transactions per Second}}. In \bibinfo{booktitle}{\emph{IEEE Int. Conf. on
  Blockchain and Cryptocurrency}}. \bibinfo{publisher}{{IEEE}},
  \bibinfo{address}{{Piscataway, NJ, USA}}, \bibinfo{pages}{455--463}.
\newblock


\bibitem[\protect\citeauthoryear{Gupta et~al\mbox{.}}{Gupta
  et~al\mbox{.}}{2018}]%
        {Gupta2018}
\bibfield{author}{\bibinfo{person}{Himanshu Gupta} {et~al\mbox{.}}}
  \bibinfo{year}{2018}\natexlab{}.
\newblock \showarticletitle{{Efficiently processing temporal queries on
  Hyperledger Fabric}}. In \bibinfo{booktitle}{\emph{Proc. - IEEE 34th Int.
  Conf. on Data Engineering, ICDE 2018}}. \bibinfo{publisher}{{IEEE}},
  \bibinfo{address}{{Piscataway, NJ, USA}}, \bibinfo{pages}{1435--1440}.
\newblock


\bibitem[\protect\citeauthoryear{Hao, Li, Dong, Fang, and Chen}{Hao
  et~al\mbox{.}}{2018}]%
        {Hao2018}
\bibfield{author}{\bibinfo{person}{Yue Hao}, \bibinfo{person}{Yi Li},
  \bibinfo{person}{Xinghua Dong}, \bibinfo{person}{Li Fang}, {and}
  \bibinfo{person}{Ping Chen}.} \bibinfo{year}{2018}\natexlab{}.
\newblock \showarticletitle{{Performance Analysis of Consensus Algorithm in
  Private Blockchain}}. In \bibinfo{booktitle}{\emph{IEEE Intelligent Vehicles
  Symposium, Proceedings}}, Vol.~\bibinfo{volume}{2018-June}.
  \bibinfo{publisher}{{IEEE}}, \bibinfo{address}{{Piscataway, NJ, USA}},
  \bibinfo{pages}{280--285}.
\newblock


\bibitem[\protect\citeauthoryear{Inagaki, Ueda, Nakaike, and Ohara}{Inagaki
  et~al\mbox{.}}{2019}]%
        {Inagaki2019}
\bibfield{author}{\bibinfo{person}{Tatsushi Inagaki}, \bibinfo{person}{Yohei
  Ueda}, \bibinfo{person}{Takuya Nakaike}, {and} \bibinfo{person}{Moriyoshi
  Ohara}.} \bibinfo{year}{2019}\natexlab{}.
\newblock \showarticletitle{{Profile-based Detection of Layered Bottlenecks}}.
  In \bibinfo{booktitle}{\emph{ICPE 2019 - Proc. of the 2019 ACM/SPEC Int.
  Conf. on Performance Engineering}}. \bibinfo{publisher}{ACM},
  \bibinfo{address}{New York, NY, USA}, \bibinfo{pages}{197--208}.
\newblock


\bibitem[\protect\citeauthoryear{Javaid, Hu, and Brebner}{Javaid
  et~al\mbox{.}}{2019}]%
        {Javaid2019}
\bibfield{author}{\bibinfo{person}{Haris Javaid}, \bibinfo{person}{Chengchen
  Hu}, {and} \bibinfo{person}{Gordon Brebner}.}
  \bibinfo{year}{2019}\natexlab{}.
\newblock \showarticletitle{{Optimizing validation phase of Hyperledger
  Fabric}}. In \bibinfo{booktitle}{\emph{IEEE Computer Society's Annual Int.
  Symp. on Modeling, Analysis, and Simulation of Computer and
  Telecommunications Systems}}, Vol.~\bibinfo{volume}{2019-Oct}.
  \bibinfo{publisher}{{IEEE} Computer Society}, \bibinfo{address}{{Piscataway,
  NJ, USA}}, \bibinfo{pages}{269--275}.
\newblock


\bibitem[\protect\citeauthoryear{Jiang et~al\mbox{.}}{Jiang
  et~al\mbox{.}}{2020}]%
        {Jiang2020}
\bibfield{author}{\bibinfo{person}{Lili Jiang} {et~al\mbox{.}}}
  \bibinfo{year}{2020}\natexlab{}.
\newblock \showarticletitle{{Performance analysis of Hyperledger Fabric
  platform: A hierarchical model approach}}.
\newblock \bibinfo{journal}{\emph{Peer-to-Peer Networking and Applications}}
  \bibinfo{volume}{13}, \bibinfo{number}{3} (\bibinfo{year}{2020}),
  \bibinfo{pages}{1014--1025}.
\newblock


\bibitem[\protect\citeauthoryear{Kuzlu et~al\mbox{.}}{Kuzlu
  et~al\mbox{.}}{2019}]%
        {Kuzlu2019}
\bibfield{author}{\bibinfo{person}{Murat Kuzlu} {et~al\mbox{.}}}
  \bibinfo{year}{2019}\natexlab{}.
\newblock \showarticletitle{{Performance analysis of a Hyperledger Fabric
  blockchain framework: Throughput, latency and scalability}}. In
  \bibinfo{booktitle}{\emph{2nd IEEE Int. Conf. on Blockchain}}.
  \bibinfo{publisher}{{IEEE}}, \bibinfo{address}{{Piscataway, NJ, USA}},
  \bibinfo{pages}{536--540}.
\newblock


\bibitem[\protect\citeauthoryear{Llanos}{Llanos}{2006}]%
        {Llanos2006}
\bibfield{author}{\bibinfo{person}{Diego~R. Llanos}.}
  \bibinfo{year}{2006}\natexlab{}.
\newblock \showarticletitle{{TPCC-UVa: An open-source TPC-C implementation for
  global performance measurement of computer systems}}.
\newblock \bibinfo{journal}{\emph{SIGMOD Record}} \bibinfo{volume}{35},
  \bibinfo{number}{4} (\bibinfo{year}{2006}), \bibinfo{pages}{6--15}.
\newblock


\bibitem[\protect\citeauthoryear{Nakaike et~al\mbox{.}}{Nakaike
  et~al\mbox{.}}{2020}]%
        {Nakaike2020}
\bibfield{author}{\bibinfo{person}{Takuya Nakaike} {et~al\mbox{.}}}
  \bibinfo{year}{2020}\natexlab{}.
\newblock \showarticletitle{{Hyperledger Fabric Performance Characterization
  and Optimization Using GoLevelDB Benchmark}}. In
  \bibinfo{booktitle}{\emph{IEEE Int. Conf. on Blockchain and Cryptocurrency}}.
  \bibinfo{publisher}{{IEEE}}, \bibinfo{address}{{Piscataway, NJ, USA}},
  \bibinfo{pages}{1--9}.
\newblock


\bibitem[\protect\citeauthoryear{Nasir, Qasse, {Abu Talib}, and Nassif}{Nasir
  et~al\mbox{.}}{2018}]%
        {Nasir2018}
\bibfield{author}{\bibinfo{person}{Qassim Nasir}, \bibinfo{person}{Ilham~A.
  Qasse}, \bibinfo{person}{Manar {Abu Talib}}, {and} \bibinfo{person}{Ali~Bou
  Nassif}.} \bibinfo{year}{2018}\natexlab{}.
\newblock \showarticletitle{{Performance analysis of hyperledger fabric
  platforms}}.
\newblock \bibinfo{journal}{\emph{Security and Communication Networks}}
  \bibinfo{volume}{2018} (\bibinfo{year}{2018}), \bibinfo{pages}{1--14}.
\newblock


\bibitem[\protect\citeauthoryear{Nguyen, Loghin, Tuan, and Dinh}{Nguyen
  et~al\mbox{.}}{2021}]%
        {Nguyen2019}
\bibfield{author}{\bibinfo{person}{Minh~Quang Nguyen},
  \bibinfo{person}{Dumitrel Loghin}, \bibinfo{person}{Tien Tuan}, {and}
  \bibinfo{person}{Anh Dinh}.} \bibinfo{year}{2021}\natexlab{}.
\newblock \showarticletitle{Understanding the Scalability of Hyperledger
  Fabric}.
\newblock \bibinfo{journal}{\emph{CoRR}}  \bibinfo{volume}{abs/2107.09886}
  (\bibinfo{year}{2021}), 10.
\newblock


\bibitem[\protect\citeauthoryear{Nguyen, Jourjon, Potop-Butucaru, and
  Thai}{Nguyen et~al\mbox{.}}{2019}]%
        {Nguyen2019b}
\bibfield{author}{\bibinfo{person}{Thanh Son~Lam Nguyen},
  \bibinfo{person}{Guillaume Jourjon}, \bibinfo{person}{Maria Potop-Butucaru},
  {and} \bibinfo{person}{Kim~Loan Thai}.} \bibinfo{year}{2019}\natexlab{}.
\newblock \showarticletitle{{Impact of network delays on Hyperledger Fabric}}.
  In \bibinfo{booktitle}{\emph{INFOCOM 2019 - IEEE Conf. on Computer
  Communications Workshops 2019}}. \bibinfo{publisher}{{IEEE}},
  \bibinfo{address}{{Piscataway, NJ, USA}}, \bibinfo{pages}{222--227}.
\newblock


\bibitem[\protect\citeauthoryear{Pongnumkul, Siripanpornchana, and
  Thajchayapong}{Pongnumkul et~al\mbox{.}}{2017}]%
        {Pongnumkul2017}
\bibfield{author}{\bibinfo{person}{Suporn Pongnumkul},
  \bibinfo{person}{Chaiyaphum Siripanpornchana}, {and}
  \bibinfo{person}{Suttipong Thajchayapong}.} \bibinfo{year}{2017}\natexlab{}.
\newblock \showarticletitle{{Performance analysis of private blockchain
  platforms in varying workloads}}. In \bibinfo{booktitle}{\emph{26th Int.
  Conf. on Computer Communications and Networks}}. \bibinfo{publisher}{{IEEE}},
  \bibinfo{address}{{Piscataway, NJ, USA}}, \bibinfo{pages}{1--6}.
\newblock


\bibitem[\protect\citeauthoryear{Shalaby et~al\mbox{.}}{Shalaby
  et~al\mbox{.}}{2020}]%
        {Shalaby2020}
\bibfield{author}{\bibinfo{person}{Salma Shalaby} {et~al\mbox{.}}}
  \bibinfo{year}{2020}\natexlab{}.
\newblock \showarticletitle{{Performance Evaluation of Hyperledger Fabric}}. In
  \bibinfo{booktitle}{\emph{IEEE Int. Conf. on Informatics, IoT, and Enabling
  Technologies}}. \bibinfo{publisher}{{IEEE}}, \bibinfo{address}{{Piscataway,
  NJ, USA}}, \bibinfo{pages}{608--613}.
\newblock


\bibitem[\protect\citeauthoryear{Sharma, Schuhknecht, Agrawal, and
  Dittrich}{Sharma et~al\mbox{.}}{2018}]%
        {Sharma2018}
\bibfield{author}{\bibinfo{person}{Ankur Sharma}, \bibinfo{person}{Felix~Martin
  Schuhknecht}, \bibinfo{person}{Divya Agrawal}, {and} \bibinfo{person}{Jens
  Dittrich}.} \bibinfo{year}{2018}\natexlab{}.
\newblock \showarticletitle{{How to Databasify a Blockchain: the Case of
  Hyperledger Fabric}}.
\newblock \bibinfo{journal}{\emph{arxiv.org}}  \bibinfo{volume}{abs/1810.13177}
  (\bibinfo{year}{2018}), 28.
\newblock


\bibitem[\protect\citeauthoryear{Sukhwani et~al\mbox{.}}{Sukhwani
  et~al\mbox{.}}{2017}]%
        {Sukhwani2017}
\bibfield{author}{\bibinfo{person}{Harish Sukhwani} {et~al\mbox{.}}}
  \bibinfo{year}{2017}\natexlab{}.
\newblock \showarticletitle{{Performance modeling of PBFT consensus process for
  permissioned blockchain network (Hyperledger Fabric)}}. In
  \bibinfo{booktitle}{\emph{IEEE Symp. on Reliable Distributed Systems}},
  Vol.~\bibinfo{volume}{2017-Sept}. \bibinfo{publisher}{{IEEE} Computer
  Society}, \bibinfo{address}{{Piscataway, NJ, USA}},
  \bibinfo{pages}{253--255}.
\newblock


\bibitem[\protect\citeauthoryear{Sukhwani et~al\mbox{.}}{Sukhwani
  et~al\mbox{.}}{2018}]%
        {Sukhwani2018}
\bibfield{author}{\bibinfo{person}{Harish Sukhwani} {et~al\mbox{.}}}
  \bibinfo{year}{2018}\natexlab{}.
\newblock \showarticletitle{{Performance modeling of Hyperledger Fabric
  (permissioned blockchain network)}}. In \bibinfo{booktitle}{\emph{17th IEEE
  Int. Symp. on Network Computing and Applications}}.
  \bibinfo{publisher}{{IEEE}}, \bibinfo{address}{{Piscataway, NJ, USA}},
  \bibinfo{pages}{1--8}.
\newblock


\bibitem[\protect\citeauthoryear{Takeshi et~al\mbox{.}}{Takeshi
  et~al\mbox{.}}{2018}]%
        {Takeshi2018}
\bibfield{author}{\bibinfo{person}{Miyamae Takeshi} {et~al\mbox{.}}}
  \bibinfo{year}{2018}\natexlab{}.
\newblock \showarticletitle{{Performance improvement of the consortium
  blockchain for financial business applications}}.
\newblock \bibinfo{journal}{\emph{Journal of Digital Banking}}
  \bibinfo{volume}{2}, \bibinfo{number}{4} (\bibinfo{year}{2018}),
  \bibinfo{pages}{369--378}.
\newblock


\bibitem[\protect\citeauthoryear{Thakkar, Nathan, and Viswanathan}{Thakkar
  et~al\mbox{.}}{2018}]%
        {Thakkar2018}
\bibfield{author}{\bibinfo{person}{Parth Thakkar}, \bibinfo{person}{Senthil
  Nathan}, {and} \bibinfo{person}{Balaji Viswanathan}.}
  \bibinfo{year}{2018}\natexlab{}.
\newblock \showarticletitle{{Performance benchmarking and optimizing
  {H}yperledger {F}abric blockchain platform}}. In
  \bibinfo{booktitle}{\emph{26th IEEE Int. Symp. on Modeling, Analysis and
  Simulation of Computer and Telecommunication Systems}}.
  \bibinfo{publisher}{{IEEE} Computer Society}, \bibinfo{address}{{Piscataway,
  NJ, USA}}, \bibinfo{pages}{264--276}.
\newblock


\bibitem[\protect\citeauthoryear{Wang and Chu}{Wang and Chu}{2020}]%
        {Wang2020}
\bibfield{author}{\bibinfo{person}{Canhui Wang} {and} \bibinfo{person}{Xiaowen
  Chu}.} \bibinfo{year}{2020}\natexlab{}.
\newblock \showarticletitle{{Performance characterization and bottleneck
  analysis of Hyperledger Fabric}}. In \bibinfo{booktitle}{\emph{Int. Conf. on
  Distributed Computing Systems}}, Vol.~\bibinfo{volume}{2020-Nov}.
  \bibinfo{publisher}{{IEEE}}, \bibinfo{address}{{Piscataway, NJ, USA}},
  \bibinfo{pages}{1281--1286}.
\newblock


\bibitem[\protect\citeauthoryear{Wang}{Wang}{2019}]%
        {Wang2019}
\bibfield{author}{\bibinfo{person}{Shuo Wang}.}
  \bibinfo{year}{2019}\natexlab{}.
\newblock \showarticletitle{{Performance Evaluation of Hyperledger Fabric with
  Malicious Behavior}}. In \bibinfo{booktitle}{\emph{Lecture Notes in Computer
  Science}}, Vol.~\bibinfo{volume}{11521}. \bibinfo{publisher}{Springer},
  \bibinfo{address}{Cham}, \bibinfo{pages}{211--219}.
\newblock


\bibitem[\protect\citeauthoryear{{World Economic Forum}}{{World Economic
  Forum}}{2019}]%
        {wef}
\bibfield{author}{\bibinfo{person}{{World Economic Forum}}.}
  \bibinfo{year}{2019}\natexlab{}.
\newblock \bibinfo{title}{{Building Value with Blockchain Technology: How to
  Evaluate Blockchain’s Benefits}}.
\newblock
\newblock
\urldef\tempurl%
\url{{https://www3.weforum.org/docs/WEF_Building_Value_with_Blockchain.pdf}}
\showURL{%
\tempurl}


\bibitem[\protect\citeauthoryear{Xu et~al\mbox{.}}{Xu et~al\mbox{.}}{2021}]%
        {Xu2021}
\bibfield{author}{\bibinfo{person}{Xiaoqiong Xu} {et~al\mbox{.}}}
  \bibinfo{year}{2021}\natexlab{}.
\newblock \showarticletitle{{Latency performance modeling and analysis for
  Hyperledger Fabric blockchain network}}.
\newblock \bibinfo{journal}{\emph{Information Processing \& Management}}
  \bibinfo{volume}{58}, \bibinfo{number}{1} (\bibinfo{year}{2021}),
  \bibinfo{pages}{102436}.
\newblock


\bibitem[\protect\citeauthoryear{Yuan, Zheng, Xiong, Zhang, and Lei}{Yuan
  et~al\mbox{.}}{2020}]%
        {Yuan2020}
\bibfield{author}{\bibinfo{person}{Pu Yuan}, \bibinfo{person}{Kan Zheng},
  \bibinfo{person}{Xiong Xiong}, \bibinfo{person}{Kuan Zhang}, {and}
  \bibinfo{person}{Lei Lei}.} \bibinfo{year}{2020}\natexlab{}.
\newblock \showarticletitle{{Performance modeling and analysis of a
  Hyperledger-based system using GSPN}}.
\newblock \bibinfo{journal}{\emph{Computer Communications}}
  \bibinfo{volume}{153} (\bibinfo{year}{2020}), \bibinfo{pages}{117--124}.
\newblock


\end{thebibliography}

\end{document}